\documentclass[journal]{IEEEtran}
\pdfoutput=1
\usepackage{ifpdf}
\usepackage{subfigure}
\usepackage{graphicx}
\DeclareGraphicsExtensions{.pdf,.jpg,.png}
\usepackage[cmex10]{amsmath}
\usepackage{amssymb}
\usepackage{array}
\usepackage{mdwmath}
\usepackage{mdwtab}
\usepackage{pslatex}
\usepackage{xcolor}
\usepackage{textcomp}
\usepackage{placeins}
\usepackage{eqparbox}
\usepackage{url}
\usepackage{stfloats}
\usepackage{multicol}
\usepackage{subfigure}
\usepackage{float}
\usepackage{algorithmicx}
\usepackage{algpseudocode}
\usepackage{amsthm}
\usepackage{comment}
\usepackage{todo}
\usepackage{color}
\usepackage{balance}

\begin{document}

\title{\vspace{-7mm}When IoT Keeps People in the Loop:\\A Path Towards a New Global Utility}

\author{
\IEEEauthorblockN{Vitaly Petrov\IEEEauthorrefmark{2}, Konstantin Mikhaylov, Dmitri Moltchanov, Sergey Andreev, Gabor Fodor,\\
Johan Torsner, Halim Yanikomeroglu, Markku Juntti, and Yevgeni Koucheryavy\vspace{-8mm}}
\thanks{V. Petrov, D. Moltchanov, S. Andreev, and Y. Koucheryavy are with the Laboratory of Electronics and Communications Engineering, Tampere University of Technology, Finland. \IEEEauthorrefmark{2}V. Petrov is the contact author: vitaly.petrov@tut.fi}
\thanks{K. Mikhaylov and M. Juntti are with the Centre for Wireless Communications, University of Oulu, Oulu, Finland.}
\thanks{H. Yanikomeroglu is with the Department of Systems and Computer Engineering, Carleton University, Canada.}
\thanks{G. Fodor is with Ericsson Research and Wireless@KTH, Sweden.}
\thanks{J. Torsner is with Ericsson Research, Finland.}
\thanks{This work was supported by Academy of Finland 6Genesis Flagship (grant 318927) and by the project TAKE-5: The 5th Evolution Take of Wireless Communication Networks, funded by Tekes. The work of V. Petrov was supported in part by a scholarship from the Nokia Foundation and in part by the HPY Research Foundation funded by Elisa. G. Fodor was sponsored by the Wireless@KTH project Naomi.}
\thanks{\copyright~2018 IEEE. Personal use of this material is permitted. Permission from IEEE must be obtained for all other uses, in any current or future media, including reprinting/republishing this material for advertising or promotional purposes, creating new collective works, for resale or redistribution to servers or lists, or reuse of any copyrighted component of this work in other works.}
}

\maketitle

\begin{abstract}
While the Internet of Things (IoT) has made significant progress along the lines of supporting individual machine-type applications, it is only recently that the importance of people as an integral component of the overall IoT infrastructure has started to be fully recognized. Several powerful concepts have emerged to facilitate this vision, whether involving the human context whenever required or directly impacting user behavior and decisions. As these become the stepping stones to develop the IoT into a novel people-centric utility, this paper outlines a path to materialize this decisive transformation. We begin by reviewing the latest progress in human-aware wireless networking, then classify the attractive human--machine applications and summarize the enabling IoT radio technologies. We continue with a unique system-level performance characterization of a representative urban IoT scenario and quantify the benefits of keeping people in the loop on various levels. Our comprehensive numerical results confirm the significant gains that have been made available with tighter user involvement, and also corroborate the development of efficient incentivization mechanisms, thereby opening the door to future commoditization of the global people-centric IoT utility.
\end{abstract}


\vspace{-5mm}
\section{Introduction and Vision}

The Internet of Things (IoT) has undergone a fundamental transformation in recent decades: departing from the legacy radio frequency identification (RFID) technology of the 1980s and the wireless sensor networks of the 1990s, which were essentially siloed ``connectivity islands'' with limited interoperability, to its present form, which is becoming increasingly interconnected and heterogeneous. Today's IoT is already a fusion of numerous networked tools and appliances, equipped with advanced computational intelligence and rich communication capabilities. More broadly, the principles of contemporary IoT overlap with and permeate many adjacent domains, including mobile and pervasive computing, as well as robotics and cyber-physical systems -- with applications ranging from smartphone-based social networking that reduces traffic and pollution in cities to mission-critical industrial automation that monitors and actuates over factory processes~\cite{Nun16}.

Beyond legacy embedded systems with constrained applicability, the emerging IoT solutions are becoming more open and integrated by adaptively combining sensors and actuators with actionable intelligence for automatic monitoring and control~\cite{editor_extra1}. However, as a multitude of interconnected and intelligent machines communicate with each other and autonomously adapt to changing contexts without user involvement, the fact that present technology is made by humans and for humans is often overlooked~\cite{reviewer2}. Indeed, modern IoT systems are still widely unaware of the human context and instead consider people to be an external and unpredictable element in their control loop. Therefore, future IoT applications will need to intimately involve humans, so that people and machines could operate synergistically. To this end, human intentions, actions, psychological and physiological states, and even emotions could be detected, inferred through sensory data, and utilized as control feedback.

This concept, which is known as human-in-the-loop (HITL), \textcolor{black}{becomes a logical next step toward truly social computing and communication in smart cities~\cite{Nun15}.} HITL opens the door to next-generation people-oriented IoT \textcolor{black}{platforms, which are aware of the people context, mobility, and even mood, thus} having more efficient and intuitive manipulation~\cite{duan17}. As users increasingly interact with such human-aware HITL systems, they may also become directly influenced by the control-loop decisions, thereby closing the loop~\cite{yang18}. In fact, people may receive control input from the system in the form of suggestions and incentives (or even penalties) to diverge from their default behavior. Accordingly, human behavior may be impacted in either space (for example, the users are encouraged to move to a less congested location) or time (for example, the users are convinced to reduce their current data demand in case the network is overloaded); this is known as the ``user-in-the-loop'' (UIL)~\cite{Sch11_1}.

With UIL, often referred to as ``layer 8'', the space-time user traffic demand may be shaped opportunistically and better matched with the actual resource supply from the people-centric wireless system. While HITL involves the user whenever human participation is desired or required and UIL extends the user's role beyond a traffic-generating and traffic-consuming black box, these trends must account for the fact that people are, in essence, walking sensor networks~\cite{Guo15}. Indeed, a wide diversity of user-owned companion devices, such as mobile phones, wearables, connected vehicles, and even drones may become an integral part of the IoT infrastructure. Hence, they can augment a broad range of applications, in which human context is useful, including traffic planning, environmental monitoring, mobile social recommendation, and public safety, among others. Therefore, we envision that -- in contrast to past concepts where the user only assists the network to receive better individual service -- future user equipment will truly merge with the IoT architecture to form a deep-fused human--machine system that efficiently utilizes the complementary nature of human and machine intelligence.

Should the IoT keep people in the loop, it has the potential to evolve into an integrated \textcolor{black}{multi-tenant} system-of-systems that may form novel, unprecedented services~\cite{Sta14}. For instance, the underlying people-centric sensor and actuator network may act as a utility -- similar to electricity and water -- creating important usable knowledge from vast amounts of data. Facilitated by this new global utility, different IoT devices \textcolor{black}{and networks} that previously had nothing to do with each other may discover and start talking to one another, thereby augmenting the current talk-by-design approaches. While the existing studies primarily focus on how the IoT can serve humans in various scenarios~\cite{editor_extra1,newnew1}, in this work we maintain that people can also assist the IoT in its daily tasks, thus closing the loop. This proposed vision renders the next-generation IoT as a genuinely multi-user, multi-tenant, and multi-application platform that can be materialized in the near future by relying on the emerging IoT radio technologies.

Following our offered vision, this article reviews and classifies the people-centric applications related to the long-range radio solutions. We then describe and compare the prominent IoT-enabling radio access technologies (RATs), namely, SIGFOX, LoRaWAN, Wi-Fi HaLow, and NarrowBand IoT. Further, we present a case study for the people-centric IoT system, which investigates how the listed RATs respond to the representative human involvement models and quantify the resultant system-wide benefits. We finally discuss attractive incentivization mechanisms for the IoT to keep its users in the loop, thus aiming for a synergy between the human-centric and the IoT-centric segments of the future Internet.

\vspace{-0.1cm}
\section{Envisaged People-Centric IoT Applications}
\label{sec:apps}

In light of the above, a contemporary perspective on the IoT expects it to soon become ``the infrastructure of the information society''. The very capable machines, \textcolor{black}{ranging from sensors to vehicles}, are already assisting humans in their daily lives. Explosive growth in the population of such connected objects leads to complex human--machine interactions that become increasingly frequent, facilitated by the HITL and UIL concepts. As these interactions intensify, many categories of people-centric IoT services emerge and are expected to be deployed over the following years:

\begin{itemize}
\item \emph{Intention- and mission-aware services}. These services primarily reflect user's current intention or desire and assist by enabling, for example, situation-aware smart commuting for pedestrians, cyclists, and drivers of scooters, trucks, and other vehicles. This group of applications can help people in a variety of use cases, from highlighting the nearest available parking space on a vehicle's head-up display in urban areas to status reporting on a display or augmented reality (AR) glasses in challenging environments, such as mines, construction sites, etc.

\item \emph{Location- and context-aware services}. Another group of services is formed by location- and context-aware applications, such as those communicating alerts from environmental sensors (for example, ``put on/take off your mask'' when entering/leaving a polluted area). Many more of these services are envisioned to be deployed in the coming years, such as identifying slippery floors and low ceilings, notifying about forgotten trash when a user is about to leave the house, and many other examples.

\item \emph{Condition- and mood-aware services}. A deeper level of IoT penetration into people's lives can be achieved by integrating city/area infrastructure with personal medical and wellness devices. For instance, dietary restrictions could be applied on a menu when ordering food or a squad leader may be advised to give a break to a worker whose blood pressure has recently gone up.
\end{itemize}

\begin{figure}[t!]
	\centering
	\includegraphics[width=1.0\columnwidth]{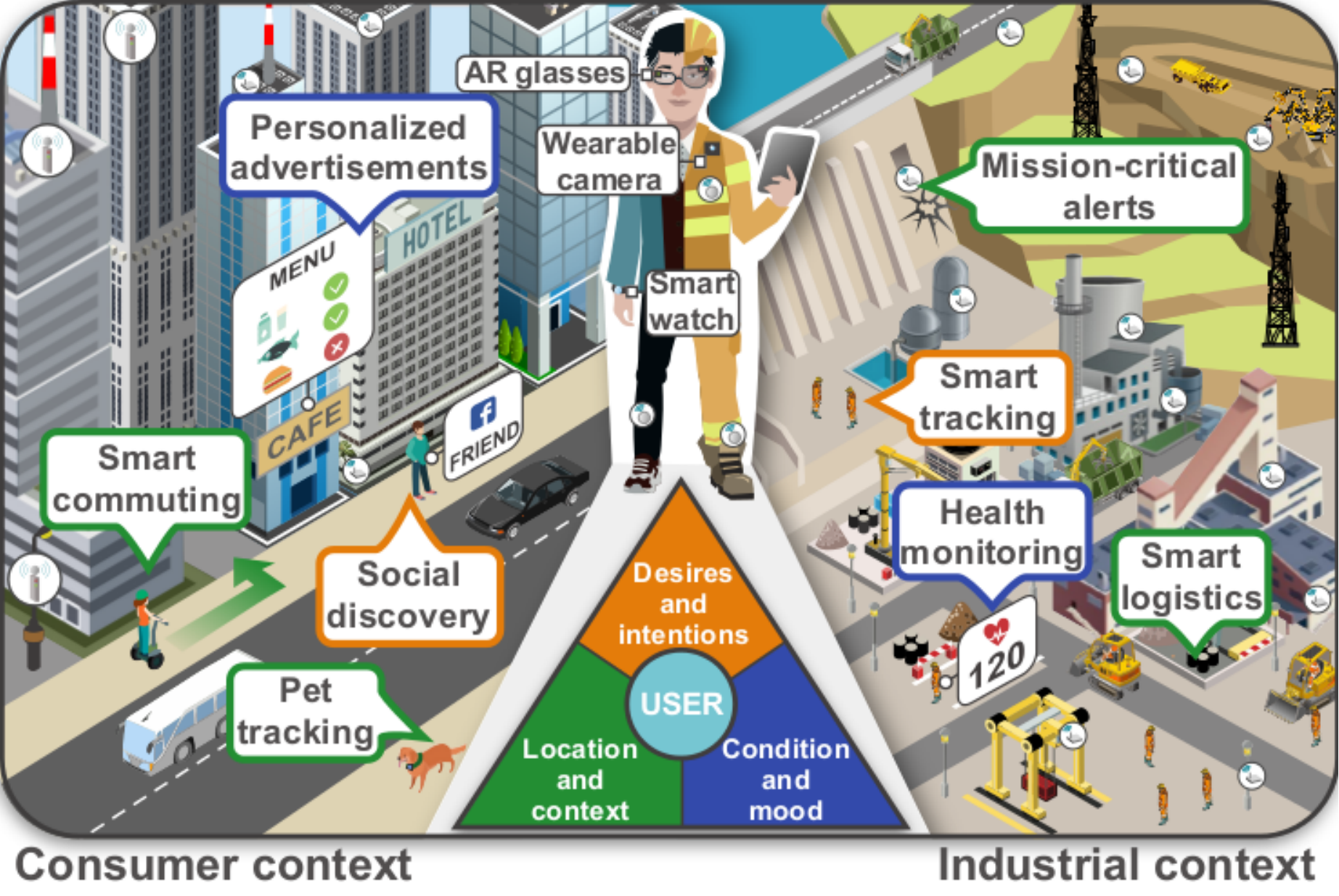}
	\caption{Consumer and industrial contexts of people-centric IoT applications.}
	\label{fig:contexts}
\vspace{-0.4cm}
\end{figure}

Summarizing the above examples of services, we note that depending on the environment the set of requirements and challenges to implement a particular application may vary considerably. To further offer a challenges-based grouping, we propose to differentiate between two major contexts: consumer and industrial (see Fig.~\ref{fig:contexts}). The former is characterized by the presence of numerous devices that are heterogeneous in terms of their communication means and ownership. Therefore, the major challenge in this context is to provide sufficient \emph{scalability} of the deployed connectivity solution. On the contrary, the latter context is more challenging in terms of maintaining communication \emph{reliability} due to more difficult propagation environments. At the same time, the system operator has more control over device population in such areas.

We continue by addressing how people-centric IoT applications are to be engineered, that is, which radio technologies need to be employed in particular scenarios and how to ensure their suitability for the target operating conditions.

\section{Review of Wireless Connectivity Options}

\begin{figure*}[t!]
	\centering
	\vspace{-8mm}
	\includegraphics[width=0.95\textwidth]{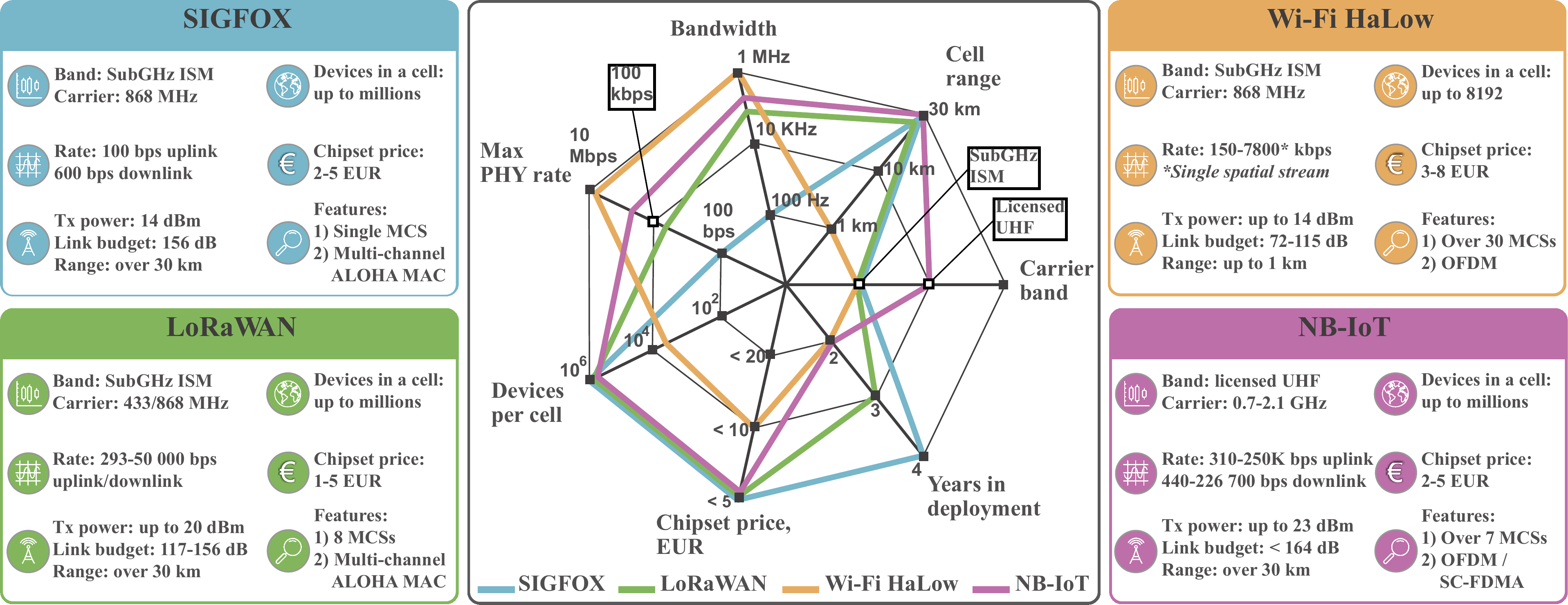}
	\vspace{-4mm}
	\caption{Key characteristics of candidate radio technologies for advanced human--machine interaction.}
	\label{fig:technologies}
\vspace{-5mm}
\end{figure*}

Inspired by the above use cases, we review the contemporary IoT radio access technologies and analyze them through the prism of their applicability for HITL applications. \textcolor{black}{Fig.~\ref{fig:technologies} brings together the major characteristics of the considered RATs.}

\subsubsection{Current technology diversity}
The demand for developing RATs that focus specifically on the needs of machines was commonly understood in the mid to late 2000s. The low power consumption, affordable cost, high communication range, and capability to handle massive deployments of infrequently transmitting devices became the major requirements for these new solutions, which can be collectively referred to as Low-Power Wide-Area Networks (LPWANs). Development of such technologies progressed in parallel within leading standardization bodies, including IEEE, ETSI, and 3GPP. 

This resulted in the sheer diversity of today's LPWAN options, which comprise a number of standardized solutions. Across this diversity, the paths taken by the technology developers differ substantially. To offer illustrative examples, we consider two emerging LPWAN solutions, namely, SIGFOX and LoRaWAN~\cite{new1}. Both technologies operate in the sub-GHz license-exempt industrial, scientific, and medical (ISM) frequency bands and employ ALOHA-based channel access with frequency hopping, which places them under severe duty cycle restrictions. Topologically, both alternatives adhere to a cellular-like structure. 

The SIGFOX solution operates with ultra-narrow band signals: $100$\,Hz using Binary Phase-Shift Keying modulation and $600$\,Hz using Gaussian Frequency Shift Keying modulation for uplink and downlink, respectively. Some limitations of this technology are harsh restrictions on the application payload of a radio frame (i.e., $12$\,bytes at most), the single modulation and coding scheme (MCS), and the limitation on the number of packets that can be sent to and from a device per day to $4$ and $140$, respectively. \textcolor{black}{Today's SIGFOX installation base exceeds 2.5 million devices.}

In contrast to SIGFOX, LoRaWAN supports multiple MCSs. The mandatory MCSs are based on the Semtech's proprietary LoRa spread spectrum modulation derived from the chirp spread spectrum modulation. The number of chirps to carry a single bit as well as the bandwidth of the channel that affects the duration of a single chirp can be adjusted to trade the on-air time for the transmission range. \textcolor{black}{There are currently several commercial deployments of LoRaWAN with the overall number of chips in excess of 5 million.}

\subsubsection{Machines talk Wi-Fi}
The path followed by the IEEE 802.11 community is substantially different. The work on IEEE 802.11ah standard (also known as Wi-Fi HaLow) has approached its final stage and early chipsets implementing this technology are already announced. The solution targets the ``high-end'' devices with demanding performance requirements in terms of throughput and latency. To this end, 802.11ah delivers a non-backward compatible communication technology in the sub-1 GHz ISM band, which is based on orthogonal frequency division multiplexing (OFDM) and features several dozens of different channel bandwidths as well as modulation and coding options. \textcolor{black}{Wi-Fi HaLow does not have any large commercial installations yet, as the technology has recently entered the testing phase.}

One of the intrigues that remain today is how close the relationship between the HaLow and the conventional Wi-Fi systems will be. That is, whether HaLow continues on its own or be merged with the traditional Wi-Fi to form dual-mode devices similar to Bluetooth Smart: for example, dual technology enabled access points or Wi-Fi HaLow access points combined with a Wi-Fi client to be used in a smartphone.

\subsubsection{3GPP goes machine}
Machine-type communication received considerable attention in 3GPP. Back in the early 2010s, 3GPP focused considerable efforts on further development of Long-Term Evolution (LTE) radio by outlining the technology named LTE-MTC or LTE-M. Addressing the need for reduced cost and energy consumption while increasing the coverage range and the number of served devices per cell, the said technology has made so far several decisive steps, while being followed by the recently standardized narrowband IoT (NB-IoT) solution.

In September 2015, following the recent activities on cellular IoT condensed in TR 45.820 document, the work on NB-IoT has officially commenced. The new Cat. NB1 devices can be integrated into today's communication networks and enable UEs with about $10$$\%$ complexity of that for Cat. 1. To achieve this goal, the bandwidth is reduced down to $180$\,kHz, orthogonal frequency-division multiple access (OFDMA) with $15$\,kHz subcarrier spacing for the downlink, frequency-division multiple access (FDMA) with $15$\,kHz subcarrier spacing and single-carrier FDMA with either $3.75$ or $15$\,kHz between the subcarriers in the uplink. NB-IoT can offer three deployment options: standalone, in-band on the LTE carrier, or in the LTE guard band. The standardization process behind NB-IoT was completed in June 2016 and the solution was included into LTE Release 13. Presently, NB-IoT and LTE-M technologies are in active commercial deployment in Americas, Europe, Asia, and Australia.

\setcounter{subsubsection}{0}

\section{Developed Evaluation Methodology}

With the aim to characterize the tentative performance gains of various user involvement mechanisms, we concentrate on a representative urban use case that may correspond to several practical IoT applications (see Fig.~\ref{fig:contexts}): smart commuting, context-aware alerts, personalized advertisements, etc. The following text summarizes the considered deployment and network topology, the proposed user involvement strategies, and the details of our conducted simulation study.

\subsubsection{Deployment parameters}
\begin{figure}[t!]
	\centering
	\includegraphics[width=0.95\columnwidth]{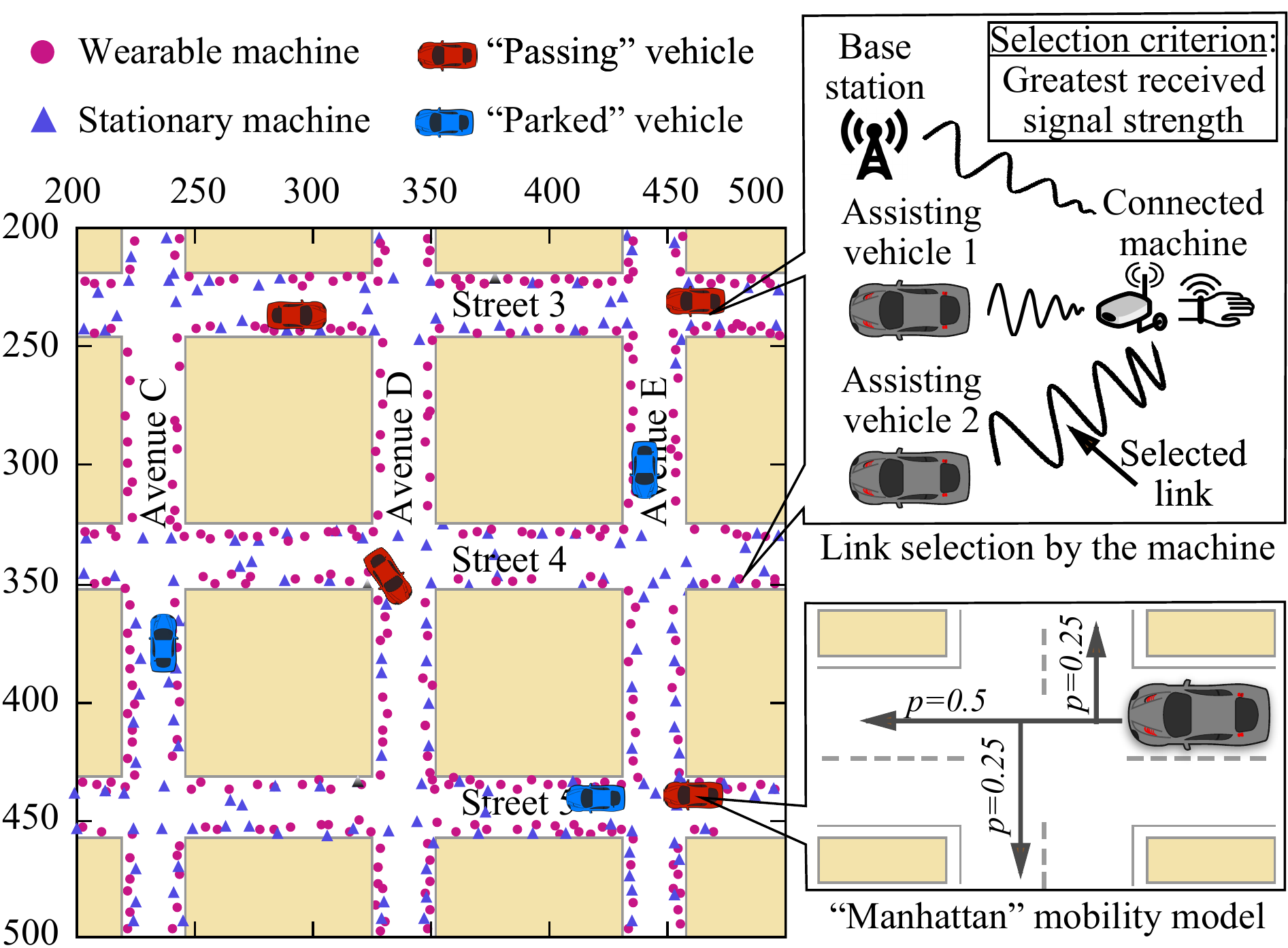}
	\vspace{-3mm}
	\caption{Part of our representative urban IoT scenario: $300$\,m~$\times$~$300$\,m out of $1$\,km~$\times$~$1$\,km area is displayed.}
	\label{fig:deployment}
\vspace{-5mm}
\end{figure}

In this work, we focus on a typical urban scenario over a $1$\,km$^2$ Manhattan grid deployment (see Fig.~\ref{fig:deployment}). A total of $100$ city blocks are modeled. We consider two types of connected machines: stationary devices that represent smart road signs, environmental sensors, etc., and mobile wearable machines that offer healthcare, fitness, and well-being functionality. Following Google StreetView data on the density of road signs and pedestrians on a weekly basis in Manhattan, New York City, USA, the number of connected machines is set to its minimum feasible value of $20$ pre-deployed devices per block and $30$ wearable devices per block. This leads to $2,000$ stationary and $3,000$ moving machines across the entire simulation scenario. All of the modeled devices are deployed uniformly over the sidewalks.

We also model vehicles that participate in intense downtown traffic, where cars are driving along the streets with the constant speed of $30$\,km/h. The random inter-vehicle distance follows an exponential distribution with the mean of $3$\,m. At the intersections, we adopt the Manhattan mobility pattern: vehicles continue in the same direction with the probability of $0.5$, while the chances to make a left or right turn are equal to $0.25$. Pedestrians carrying wearable machines follow a similar mobility pattern with the speed of $5$\,km/h. To eliminate border effects \textcolor{black}{and maintain uniform density of mobile objects (namely, moving machines and assisting vehicles),} a wraparound mechanism is implemented, such that any mobile device leaving the area of interest on a side immediately appears on the opposite side of the map.

\begin{table}[t]
\centering
\vspace{-2mm}
\caption{Simulation setup and parameters}
\vspace{-2mm}
\begin{tabular}{l|l}
\hline
\textbf{Parameter} & \textbf{Value}\\
\hline
Message payload (stationary machine) & $10$ B\\
Message payload (moving machine) & $100$ B\\
Inter-arrival time (stationary machine) & $5$\,s\\
Inter-arrival time (moving machine) & $60$\,s\\
City block size & $80$\,m\\
Street width & $25$\,m\\
Scenario size ($100$ blocks) & $1050$\,m $\times$ $1050$\,m\\
\hline
Total number of stationary machines & $2,000$ / scenario\\
Total number of moving machines & $3,000$ / scenario\\
Base station height & $10$ m\\
Car roof height & $1.5$ m\\
Deployment height of stationary machines & Uniform in $[0;10)$ m\\
Deployment height of moving machines & $1.5$\,m\\
\hline
Number of vehicles & $1,000$ / scenario\\
Number of pedestrians & $3,000$ / scenario\\
Speed of vehicles & $30$~km/h\\
Speed of pedestrians & $5$~km/h\\
Average inter-vehicle distance & $3$~m\\
Inter-vehicle distance distribution & Exponential\\
Mobility pattern of vehicles and pedestrians & ``Manhattan''\\
\textcolor{black}{Association rule} & \textcolor{black}{Max received power}\\
\hline
\end{tabular}
\label{tab:parameters}
\vspace{-0.5cm}
\end{table}

\subsubsection{Simulation details}
The reported performance assessment has been conducted with our custom-made system-level simulator, named WINTERsim, which has been extensively utilized recently for studying various IoT scenarios~\cite{M2M_paper_dohler}. 
In the present study, we focus primarily on the uplink IoT traffic by modeling connectivity between a number of machines and the base station (BS). To fairly compare the behavior of all the four considered RATs (i.e., to avoid overloading SIGFOX), the data transmissions are assumed to be regular and infrequent. More specifically, stationary machines communicate $10$\,B messages every $5$\,s, while moving devices send a $100$\,B update every minute, which corresponds to the typical sensing-based applications (e.g., assessing user's physical condition with a medical sensor). 

From the connectivity perspective, machines communicate with their serving BS by default (termed \emph{baseline} connectivity). Alternatively, if a better signal level is available, the devices may also connect to one of the femtocell relay stations deployed on the mobile vehicles across the tracking area (termed \emph{assisted} connectivity). To characterize the effects of involving user-owned relaying cars for the ``average'' IoT device, ``median-quality'' connections to the BS are of particular interest. Therefore, in our scenario the BS is located at a certain distance from the center of the area to ensure the baseline average signal-to-interference-plus-noise ratio (SINR) of $10$\,dB. Since the properties of the considered LPWAN RATs vary significantly, four different simulation setups have been considered (one per RAT), with the distances to the BS ranging from $520$\,m for Wi-Fi HaLow to $12$\,km for LoRaWAN. For the sake of better accuracy, the numerical results in each of these evaluations have been averaged over $100$ independent simulation rounds.

Certain additional assumptions have been made to implement the LPWAN technologies of interest. Particularly, the hard message limitation for SIGFOX was relaxed. Then, LoRaWAN was restricted to operate exclusively over the $868$\,MHz band. The bandwidth of Wi-Fi HaLow, in its turn, was set to $1$\,MHz by disregarding the $2$\,MHz and wider bandwidth options that offer higher data rates for the cost of reduced coverage. Finally, only seven key MCSs were supported for NB-IoT. The remaining radio technology related parameters were adopted from Fig.~\ref{fig:technologies}, while other important settings are summarized in Table~\ref{tab:parameters}.

In our subsequent evaluation, we focus on the two key metrics: (i) SINR at the receiver and (ii) energy efficiency of machines, which is defined as the amount of data that has been reliably delivered from the IoT device to the BS in relation to the total energy consumed by the machine's radio interface.

\subsubsection{Degrees of user involvement}\label{sec:user_involvement}
As relaying vehicles in our IoT scenario are not owned by the operator but by the private users, the assisted connectivity option requires certain levels of user involvement. In this article, we study two different types of such engagement in the characteristic urban IoT deployment:

$\bullet$ \underline{\emph{User involvement Type 1}}. In this case, some of the driving vehicles share their communication capabilities and, if in close proximity, can opportunistically forward the IoT traffic from the machines to the application server. The deployed machines compare the connection quality to the BS with that to the nearest vehicle that is willing to assist, and transmit their subsequent update over a better channel. \textcolor{black}{In particular, the signal strength of the received beacon has been considered here as the selection criterion.} In its turn, the assisting vehicle relays thus received IoT data to the application server over its on-board communication equipment.

\begin{figure*}[t!]
\centering
\vspace{-5mm}
\includegraphics[width=0.95\textwidth]{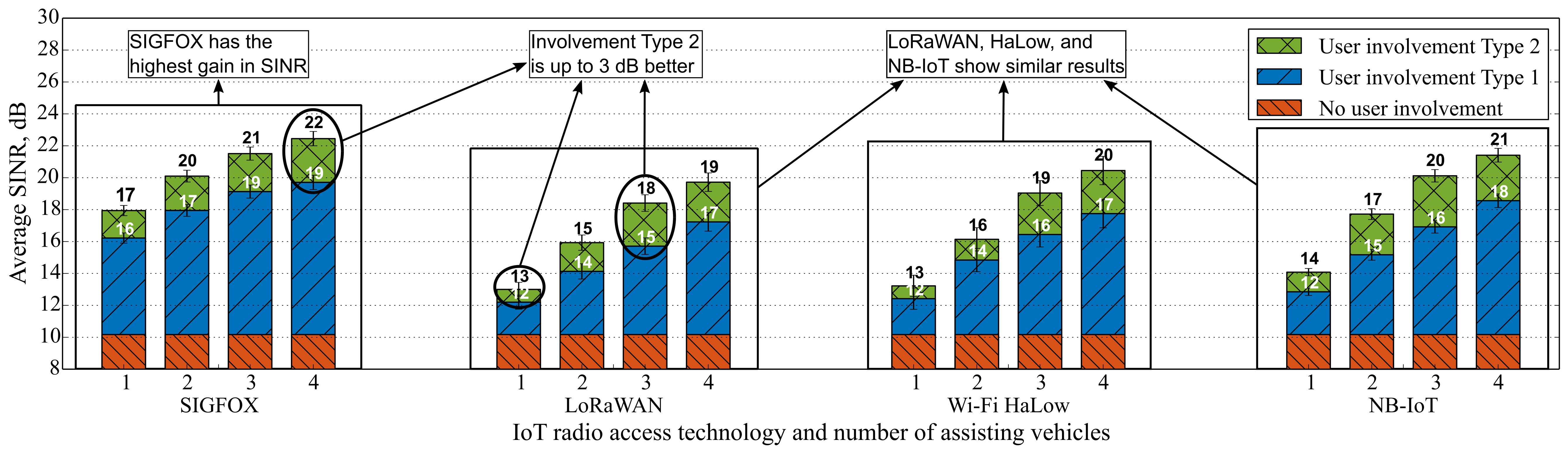}
\vspace{-5mm}
\caption{Average SINR values with various user involvement types.}
\label{fig:sinr}
\vspace{-3mm}
\end{figure*}

$\bullet$ \underline{\emph{User involvement Type 2}}. As the user involvement Type 1 does not affect the vehicle mobility patterns, it can only offer opportunistic gains. On the contrary, the user involvement Type 2 suggests the car owner to temporarily park the vehicle close to the cluster of machines suffering from inadequate connection quality. For instance, the car owner may be offered as a reward a discounted parking permit in the said location (or free charging time in case of an electric vehicle). In this case, conforming vehicles are placed next to the centers of device clusters and start continuously forwarding the IoT data from the machines to the application server. Accordingly, a link with a better quality can be made available to the machines for longer intervals of time. However, the user involvement Type 2 may require humans to deviate from their intended mobility patterns and the corresponding sources of motivation should thus be provided.

The achievable performance gains in terms of machine's energy efficiency for both types of user involvement are reported in subsection~\ref{sec:numerical}, while a discussion on the nature and the origins of user involvement is offered in subsection~\ref{sec:incentivisation}.

\section{Benefits and Nature of User Involvement}

\subsection{Achievable Performance Gains}
\label{sec:numerical}

To ease further exposition, we only account for the proportion of data transmitted by the machines. To this aim, we reasonably assume that the wireless connections from the privately-owned vehicles (which are neither battery- nor power-constrained) to the BS are reliable enough to guarantee the delivery of the relayed IoT data.

\begin{figure*}[t!]
\centering
\vspace{-5mm}
\includegraphics[width=0.95\textwidth]{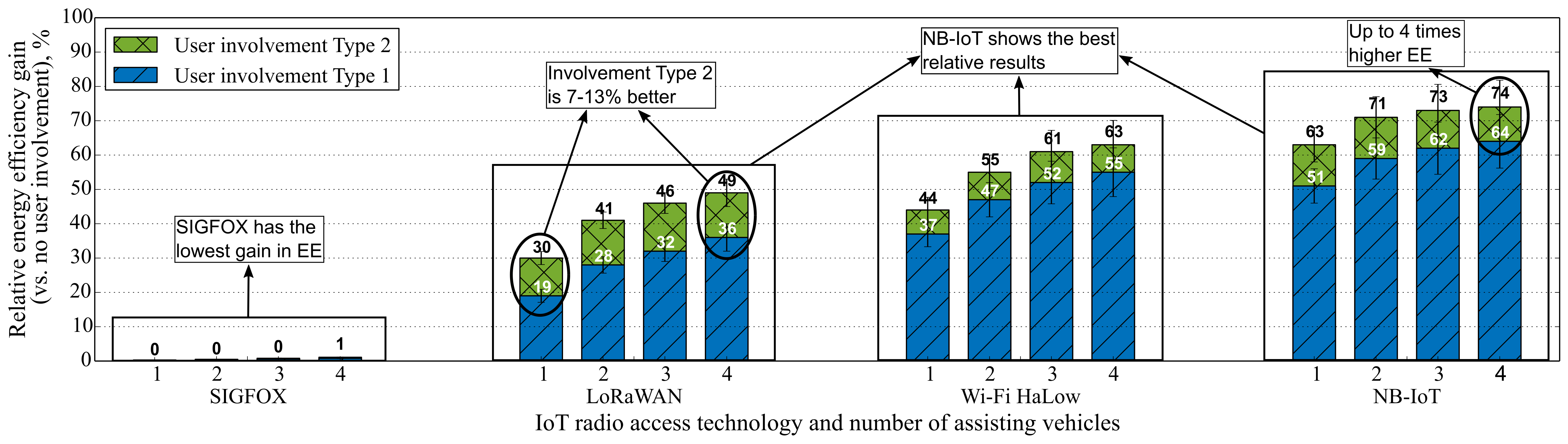}
\vspace{-5mm}
\caption{Energy-efficiency (EE) gains with various user involvement types.}
\label{fig:ee}
\vspace{-5mm}
\end{figure*}

First, Fig.~\ref{fig:sinr} reports on the average levels of SINR at the receiver (either the BS or the assisting vehicle, depending on the machine's current connectivity option). We learn that for all the IoT technologies in question the considered user involvement options notably outperform the baseline alternative. In particular, noticeable SINR gains are achieved already with a few assisting vehicles. The SINR gains for the SIGFOX solution are the highest, while the other three LPWAN protocols behave similarly as the degree of human assistance increases. The explanation behind these results is in the fact that SIGFOX operates in a full-power regime with a single data rate, while other RATs allow the machines to dynamically select their transmit power and MCS. The results for the user involvement Type 2 are, on average, several dB higher than the corresponding numbers for the user involvement Type 1, which confirms that careful positioning of assisting nodes is preferred over their opportunistic placement.

Further, Fig.~\ref{fig:ee} illustrates the impact of the studied user involvement types on the average energy efficiency of the communicating machines. This parameter offers insights into how far the battery life could be extended and, consequently, how much the operating costs could be reduced. In the figure, we first observe a marginal impact of user involvement on the energy efficiency of SIGFOX. Albeit substantial improvements in the SINR values are confirmed by Fig.~\ref{fig:sinr}, the single transmit power level in SIGFOX limits the potential benefits of the assisting vehicles. The observed gain of only several percent is due to a slightly increased level of reliability.

On the contrary, the alternative three RATs demonstrate considerable energy efficiency improvements (e.g., up to $4$ times higher energy efficiency for NB-IoT with user involvement Type 2). The observed growth is not only due to better channel conditions (which are similar for the said technologies in the same scenario), but also due to the flexibility of the IoT RATs themselves (specifically, more freedom in the MCS and transmit power selection). \emph{In summary, the considered ways of user involvement may lead to notable improvements in the IoT service reliability (due to higher SINR) as well as in the energy efficiency of networked machines. Hence, people in the loop assist the operator in reducing both the capital investments and the operating costs.} We now discuss several approaches to materializing these tentative gains.

\vspace{-2mm}
\subsection{User Incentivization Options}
\label{sec:incentivisation}

A crucial component of the considered system operation is conditional on user's willingness to share own resources while helping improve the IoT network performance. Even though the concept of user involvement for resource provisioning has been discussed for years in the context of various mobile systems, these mechanisms have not been implemented widely as of yet. This is in part owing to conservative human nature that resists new models of service provisioning, especially when the process requires involving own resources without immediately perceived benefits.

We expect that appropriate user incentivization mechanisms need to be natively integrated into the envisioned IoT infrastructure. In systems with regular topologies and central control -- that emerge primarily in the industrial context -- there always remains an opportunity to enable resource sharing by design. On the contrary, strictly enforcing resource sharing in consumer scenarios may lead to customer dissatisfaction and therefore clever incentivization mechanisms are required.

To develop appropriate incentivization schemes for the consumer context, one has to mediate between dissimilar interests of at least five major stakeholders: (i) owners of the pre-deployed machines; (ii) owners of the wearable machines; (iii) owners of the assisting vehicles; (iv) car manufacturers, and (v) system operators. \textcolor{black}{Furthermore, in order to deploy the discussed mechanisms in practice}, the operators need to carefully ``moderate the dialog'' between all of these parties, where the most non-trivial aspect is to engage the owners of vehicles.

In order to make it happen, a service provider has to offer (i) a type of compensation for the car owners and (ii) a corresponding billing mechanism. For the former, there is a set of options including those directly related to the operator--user relationships, such as decreased monthly rates and/or extended subscription periods, as well as priority in service, guaranteed rate during congestion, etc.

Further, a billing mechanism is a necessary component of the incentive-driven services. The set of requirements imposed on its choice includes robustness to a loss of control connection, high levels of security during authorization and encryption of all the exchanged billing-specific information, operational accuracy, as well as simplicity. There are two fundamental paths to implementing an efficient billing algorithm. 

In a fully centralized system, the operators become responsible for all the phases, such as collection, storage, and interpretation of information on the individual contributions by the users. To alleviate the cases of false notification by malicious participants, each session has to be authorized and data needs to be collected from both endpoints. \textcolor{black}{In case of a multi-tenant deployment, a dedicated entity needs to be empowered to account, store, and distribute the rewards collected by the stakeholders~\cite{reviewer1}.}

An alternative to the above is a network-assisted decentralized billing solution~\cite{ebit}. There are numerous practical contexts, where decentralized incentivization mechanisms have been deployed successfully. A famous example is peer-to-peer file sharing and streaming services. There, reciprocity-based mechanisms where the peers maintain and share the information about other peers' contributions remain a popular method. In such schemes, network assistance is still required for authentication purposes, but the actual billing process could be made decentralized.

\vspace{-0.2cm}
\section{Conclusions}
The unprecedented proliferation of the interconnected autonomous machines drastically increases the intensity and the depth of human--machine interactions. On the one hand, this introduces a wide range of novel challenges with respect to how the individual procedures, technologies, as well as the overall working services and applications should be engineered -- mindful of the unique capabilities and limitations of humans and machines alike. On the other hand, this brings along an excellent opportunity to jointly engage both sides to take advantage of rich mutual collaboration and shield any weaknesses with each other's strengths. In this article, we have confirmed that this remains valid for people as well as for connected machines.

Our novel analysis of an illustrative consumer IoT scenario for different user involvement levels and over four perspective radio technologies demonstrates that even moderate human assistance makes a decisive difference and unlocks significant energy savings for networked machines. Importantly, our offered numerical results suggest that while each of the addressed wireless solutions does benefit from keeping people in the loop, the actual gains for every system profile depend on the flexibility of the underlying access protocol. At the same time, the costs associated with involving the human context may include certain incentivization-related expenditures. These \textcolor{black}{may vary a lot in the envisaged heterogeneous, multi-tenant deployments, which} calls for careful planning of user engagement in future people-centric IoT infrastructure. Ultimately, the rapidly maturing human-aware IoT ecosystem may become a new global utility, thus ``disappearing'' in the ``fabric of everyday life''~\cite{Cor10}, while enabling a plethora of next-generation applications and services.

\vspace{-0.2cm}
\section*{Acknowledgment}
The authors are particularly grateful to Yngve Sel\'en (Wireless Access Networks Department, Ericsson Research) for the insightful comments that allowed to improve this paper.

\balance
\section*{Authors' Biographies}

\textbf{Vitaly Petrov} (vitaly.petrov@tut.fi) is a PhD candidate at the Laboratory of Electronics and Communications Engineering at Tampere University of Technology (TUT), Finland. He received the Specialist degree (2011) from SUAI University, St. Petersburg, Russia, as well as the M.Sc. degree (2014) from TUT. He was a Visiting Scholar with Georgia Institute of Technology, Atlanta, USA, in 2014. Vitaly (co-)authored more than 30 published research works on terahertz band communications, Internet-of-Things, nanonetworks, cryptology, and network security.

\textbf{Konstantin Mikhaylov} (konstantin.mikhaylov@oulu.fi) is a post-doctoral research fellow at the Centre of Wireless Communications, University of Oulu. He received his B.Sc. (2006) and M.Sc. (2008) degrees in electrical engineering from St. Petersburg State Polytechnical University, and Ph.D. degree from the University of Oulu in 2018. His has (co-)authored more than 60 published research papers focusing on the energy efficient radio access technologies for massive IoT, and IoT device design and development.

\textbf{Dmitri Moltchanov} (dmitri.moltchanov@tut.fi) is a senior research scientist at Tampere University of Technology (TUT). He received his M.Sc. and Cand.Sc. degrees from Saint Petersburg State University of Telecommunications in 2000 and 2002, respectively, and his Ph.D. degree from TUT in 2006. His research interests include performance evaluation and optimization issues in IP networks, Internet traffic dynamics, and traffic localization in P2P networks.

\textbf{Sergey Andreev} (sergey.andreev@tut.fi) is a Senior Research Scientist in the Laboratory of Electronics and Communications Engineering at Tampere University of Technology, Finland. He received the Specialist degree (2006) and the Cand.Sc. degree (2009) both from St. Petersburg State University of Aerospace Instrumentation, St. Petersburg, Russia, as well as the Ph.D. degree (2012) from Tampere University of Technology. Sergey (co-)authored more than 100 published research works on wireless communications, energy efficiency, and heterogeneous networking.

\textbf{Gabor Fodor} (gabor.fodor@ericsson.com) received the M.Sc. and Ph.D. degrees in electrical engineering from the Budapest University of Technology and Economics in 1988 and 1998 respectively. He is currently a master researcher at Ericsson Research and an adjunct professor at the KTH Royal Institute of Technology, Stockholm, Sweden. He was a co-recipient of the IEEE Communications Society Stephen O. Rice prize in 2018. He is serving as an Editor of the IEEE Transactions on Wireless Communications.

\textbf{Johan Torsner} (johan.torsner@ericsson.com) is a research manager in Ericsson Research and is currently leading Ericsson's research activities in Finland. He joined Ericsson in 1998, and has held several positions within research and R\&D. He has been deeply involved in the development and standardization of 3G and 4G systems, and has led over 100 patent applications. His current research interests include 4G evolution, 5G, and machine-type communication. 

\textbf{Halim Yanikomeroglu} (halim@sce.carleton.ca) is a full professor in the Department of Systems and Computer Engineering at Carleton University. His research interests cover many aspects of wireless technologies with special emphasis on cellular networks. His collaborative research with industry has resulted in 24 granted patents (plus about a dozen applied). He is a Fellow of the IEEE, a Distinguished Lecturer for the IEEE Communications Society, and a Distinguished Speaker for the IEEE Vehicular Technology Society.

\textbf{Markku Juntti} (markku.juntti@oulu.fi)received a Dr.Sc. (Tech.) degree in electrical engineering from the University of Oulu, Finland, in 1997. He has been with the University of Oulu since 1992. In 1994--1995 he visited Rice University, Houston, Texas. He has been a professor of telecommunications at the University of Oulu since 2000. His research interests include communication and information theory, signal processing for wireless communication systems, and their application in wireless communication system design.

\textbf{Yevgeni Koucheryavy} (evgeni.kucheryavy@tut.fi) is a Full Professor at the Laboratory of Electronics and Communications Engineering of Tampere University of Technology (TUT), Finland. He received his Ph.D. degree (2004) from TUT. He is the author of numerous publications in the field of advanced wired and wireless networking and communications. He is Associate Technical Editor of IEEE Communications Magazine and Editor of IEEE Communications Surveys and Tutorials.

\end{document}